\documentclass[
 reprint,
 amsmath,amssymb,
 aps,longbibliography
]{revtex4-1}

\usepackage{graphicx}
\usepackage{dcolumn}
\usepackage{bm}
\usepackage{physics}
\usepackage{multirow}
\usepackage{xcolor}
\usepackage{verbatim}

\begin{document}

\preprint{APS/123-QED}

\title{Loss-tolerant transmission of multiphoton-qubit information via hybrid entanglement}

\author{Seongjeon Choi}
\author{Seok-Hyung Lee}
\author{Hyunseok Jeong}
\affiliation{Department of Physics and Astronomy, Seoul National University, Seoul 08826, Republic of Korea
}

\date{\today}

\begin{abstract}
It was shown that using multiphoton qubits, a nearly deterministic Bell-state measurement can be performed with linear optics and on-off photodetectors [Phys.~Rev.~Lett.~114,~113603~(2015)]. However, multiphoton qubits are generally more fragile than single-photon qubits under a lossy environment. In this paper, we propose and analyze a scheme to teleport multiphoton-qubit information using hybrid entanglement with a loss-tolerant carrier qubit. 
We consider three candidates for the carrier qubit: a coherent-state qubit, a single-photon polarization qubit, and a vacuum-and-single-photon qubit. We find that teleportation with the vacuum-and-single-photon qubit tolerates about 10 times greater photon losses than with the multiphoton qubit of the photon number $N \geq 4$ in the high fidelity regime ($F\geq 90\%$). The coherent-state qubit encoding may be even better than the vacuum-and-single-photon qubit  as the carrier when its amplitude is as small as $\alpha<0.78$. We further point out that the fidelity of the teleported state by our scheme is determined by loss in the carrier qubit  while
the success probability depends on loss only in the multiphoton qubit to be teleported. Our study implies that the hybrid architecture may complement the weaknesses of each qubit encoding. 
\end{abstract}

\maketitle

\section{\label{sec:intro}Introduction}

Photonic qubits are particularly useful for quantum information transfer over a long distance.
There are several different ways to encode qubit information in traveling light fields.
Probably the most well-known type uses the horizontal and vertical polarizations of a single photon (PSP), $\ket{H}$ and $\ket{V}$ \cite{Knill01, Dodd03}, which is often called ``dual-rail encoding.'' 
Another method is to utilize the vacuum and  the single-photon (VSP) states, $\ket{0}$ and $\ket{1}$, called ``single-rail encoding,'' with its own merit \cite{Lee00, Lund02}.
Not only restricted to the discrete qubit encoding, one can alternatively utilize continuous-variable-based qubit encodings such as one with two coherent states with opposite phases, $\ket{\pm\alpha}$,  where  $\pm\alpha$ are coherent amplitudes. This approach enables one to perform nearly deterministic Bell-state measurements \cite{Jeong01, Jeong01b} and efficient gate operations for quantum computing \cite{Jeong02, Ralph03, Lund08}. Hybrid architectures of these qubit encodings have also been explored to combine their advantages \cite{Park12, Lee13, Kwon13, Sheng13, Jeong14, Jeong15, Andersen15, Kim16, Lim16}. 

Recently, Lee \textit{et al.}~suggested multiphoton encoding with the horizontal and vertical polarizations of $N$ photons, $\{ \ket{H}^{\otimes N} = \bigotimes_{i = 1}^N \ket{H}_i, \ket{V}^{\otimes N} = \bigotimes_{i = 1}^N \ket{V}_i \}$, in order to overcome the limitation of Bell-state measurement using linear optics \cite{Lee15}. Using linear optics with single-photon qubits, only two among four Bell states can be discriminated, and the success probability of Bell measurement is generally limited to $1/2$ \cite{Lutkenhaus99, Calsamiglia01}. This affects the success probabilities of gate operations for linear optics quantum computing \cite{Knill01} depending on the gate teleportation scheme \cite{Gottesman99}, which is an obstacle against the implementation of scalable optical quantum computation. There are a number of proposals to circumvent this limitation using ancillary states or operations \cite{Grice11, Zaidi13, Ewert14, Kilmer19}, coherent-state qubits \cite{Jeong01}, hybrid qubits \cite{Lee13}, and multiphoton qubits \cite{Lee15}. Among them, the multiphoton encoding achieves a nearly deterministic Bell-state measurement with an average success probability $1- 2^{-N}$, where $N$ is the number of photons per qubit \cite{Lee15}.  Recently, it was shown that the multiphoton encoding is particularly advantageous for quantum communication \cite{Lee19}.

A multiphoton qubit is generally in the form of the Greenberger-Horne-Zeilinger~(GHZ) state, i.e.,~$\ket{\psi} = a \ket{H}^{\otimes N} + b \ket{V}^{\otimes N}$. The GHZ-type state is fragile under photon loss \cite{Simon02, Dur02}, and this makes it hard to transmit  quantum information over long distance via the multiphoton qubit. One solution to this problem is to use the parity encoding with quantum error correction that corrects photon loss errors \cite{Munro12, Muralidharan14, Ewert17, Lee19}. However, such a qubit encoding has a complex structure making it generally hard to generate the desired logical qubit and Bell states (the scheme and its success rate are discussed in Ref.~\cite{Ewert17}).

In this paper, we suggest and investigate a teleportation scheme via hybrid entanglement between a multiphoton qubit and another type of optical qubit serving as a loss-tolerant carrier. Our strategy is to send the loss-tolerant carrier qubit through the noisy environment while storing the multiphoton qubit as intact as one can. A similar  type of approach was used for loss-tolerant quantum relay for a coherent-state qubit via an asymmetric entangled coherent state~\cite{Neergaard-Nielsen13}. We consider three types of carrier qubits: a coherent-state qubit, a PSP qubit, and a VSP qubit. We investigate quantum fidelities for the output states and success probabilities of quantum teleportation under photon loss. 
The success probability of the Bell measurement is affected only by photon loss on the multiphoton qubit but the fidelity is determined by properties on the carrier qubit. It shall be shown that any choice among the three candidates can improve the fidelity.
We mainly consider the photon number for a multiphoton qubit as $N=4$ which was identified as the optimal number for fault-tolerant quantum computing using the multiphoton qubits, the seven-qubit Steane code and the telecorrection protocol \cite{Lee15, Lee15_2}.
Remarkbly, the VSP qubit in hybrid entanglement serves as a highly efficient carrier showing about 10 times better tolerance to photon loss than the direct transmission of the multiphoton qubit 
when the fidelity is larger than 0.9. The coherent-state qubit encoding can be even better than the VSP qubit  as the carrier when its amplitude is as small as $\alpha<0.78$.
Our study may be useful for designing and building up loss-tolerant quantum communication networks.


\section{\label{sec:loss_env}Photon-loss model}

We describe the environment with the photon-loss model by the Master equation under the Born-Markov approximation with the zero-temperature \cite{Phoenix90}:
\begin{align}\label{eq:masterequation}
    \frac{\partial \rho}{\partial \tau} =  \gamma \sum_{i=1}^{N}  \left( \hat{a}_i \rho \hat{a}_i^\dagger -\frac{1}{2}\hat{a}_i^\dagger \hat{a}_i \rho -\frac{1}{2}\rho \hat{a}_i^\dagger \hat{a}_i \right)
\end{align}
where $\hat{a}_i$($\hat{a}_i^\dagger$) represents the annihilation (creation) operator of mode $i$ and $\gamma$ is the decay constant determined by the coupling strength of the system and the environment. 
This evolution of a density operator is equivalently described by the beam-splitter model where each input mode is independently mixed with a vacuum state at a beam splitter with transmittance $t=e^{-\gamma \tau/2}$ and reflectance $r = \sqrt{1-t^2}$ \cite{Leonhardt93}:
\begin{align} \label{eq:beamsplitter}
    \begin{pmatrix}
    \hat{a} \\ \hat{b}
    \end{pmatrix} 
    \rightarrow
    \begin{pmatrix}
        \hat{a}' \\ \hat{b}'
    \end{pmatrix}
    = 
    \begin{pmatrix}
    t & - r \\
    r & t
    \end{pmatrix}
    \begin{pmatrix}
    \hat{a} \\ \hat{b}
    \end{pmatrix}.
\end{align}
where $\hat{a}$($\hat{b}$) is the annihilation operator on system (ancillary) mode.  The output state is then obtained by tracing out the ancillary modes. Considering the evolution of single photon state $\dyad{1} \rightarrow t^2\dyad{1} + r^2 \dyad{0}$, we refer to the sqaure of the reflactance $r^2$ as the photon-loss rate $\eta$.

\begin{figure}
    \centering
    \includegraphics[width=\linewidth]{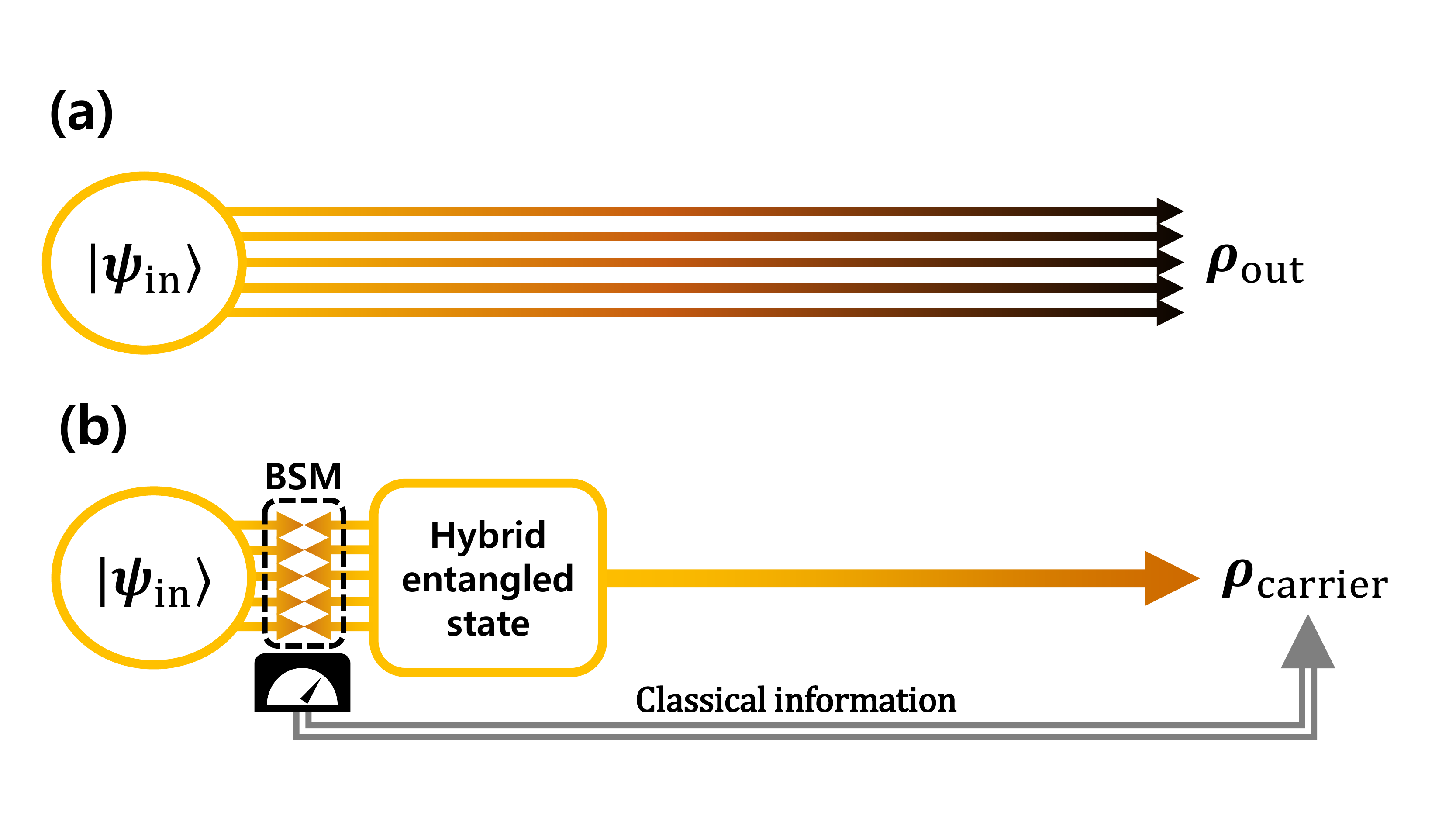}
    \caption{Schematic of quantum information transmission of a multiphoton qubit. 
    (a) A multiphoton qubit $\ket{\psi_\mathrm{in}}$ is directly transmitted.
     (b) The qubit encoding is changed to the carrier qubit by teleportation with a hybrid entangled state. 
     The classical information from the Bell-state measurement (BSM) is transmitted via a classical channel.
     }
    \label{fig:overall_shceme}
\end{figure}

\section{\label{sec:dir_transmission}Direct transmission}

Suppose that we directly transmit a multiphoton qubit of $N$ photons $\ket{\psi_\mathrm{in}}=a \ket{H}^{\otimes N}+b\ket{V}^{\otimes N}$ over a lossy environment. The output qubit of the transmission is obtained using Eq.~(\ref{eq:beamsplitter}) as
\begin{align*}
    \rho_\mathrm{out}(t) =&  \abs{a}^2 \qty[t^2 \dyad{H} + (1 - t^2) \dyad{0}]^{\otimes N} \nonumber\\
    &+ \abs{b}^2 \qty[t^2 \dyad{V} + (1 - t^2) \dyad{0}]^{\otimes N} \nonumber\\
    &+ t^{2N} [ a b^*(\dyad{H}{V})^{\otimes N} + \mathrm{H.c.}] \nonumber\\
    =& t^{2N} \dyad{\psi_\mathrm{in}} + (1-t^{2N})\rho^\mathrm{loss},
\end{align*}
where 
\begin{align}\label{eq:dir_loss}
    \rho^\mathrm{loss} = &\sum_{k=1}^{N}  (t^{2})^{N-k}(1-t^2)^{k} \sum_{\mathcal{P}\in \mathrm{Perm}(N,k)}\nonumber\\
    &\Big\{\abs{a}^2\mathcal{P}[(\dyad{H})^{\otimes N-k}\otimes (\dyad{0})^{\otimes k}]\nonumber\\
    &+ \abs{b}^2 \mathcal{P}[(\dyad{V})^{\otimes N-k}\otimes (\dyad{0})^{\otimes k}]\Big\}
\end{align} 
is the loss term with one or more photons lost. We denote $\mathrm{Perm}(N,k)$ as the set of permutations of tensor products with the number of elements $\genfrac{(}{)}{0pt}{}{N}{k}$, which represents the cases that photons in $k$ modes within the total $N$ modes are lost and photons in $N-k$ modes remain in the polarization state. It is straightforward to see that $\rho^\mathrm{loss}$ is orthogonal to $\ket{\psi_\mathrm{in}}$. 
 The quality of the output state is measured by fidelity $F$ between the input and output states that is defined as $F(t) = \bra{\psi_\mathrm{in}}\rho_\mathrm{out}(t)\ket{\psi_\mathrm{in}}$. The fidelity for the direct transmission is then obtained as
\begin{align*}
F^\mathrm{dir} =t^{2N}=(1-\eta)^N.
\end{align*}
This shows that the multiphoton qubit becomes more fragile when photon number $N$ per qubit becomes larger. 
Although the success probability of the Bell-state measurement using multiphoton qubits approaches the unity as $N$ gets larger \cite{Lee15},
this fragility may be  a weak point of the multiphoton encoding when it is applied to quantum information transfer.

\section{\label{sec:hybird}Teleportation with hybrid entanglement}

In our scheme, a hybrid entangled state between a multiphoton qubit  and a carrier qubit is used as the quantum channel, where 
 the carrier qubit is loss-tolerant compared to the multiphoton qubit.
In what follows, we examine a coherent-state qubit, a PSP qubit  and a VSP qubit as candidates for   the carrier qubit.

\subsection{\label{subsec:loss}Loss on hybrid entangled states}

For the teleportation between two different types of qubits, the sender and the receiver need to share a hybrid entangled state between a multiphoton qubit and a carrier qubit. The entangled state for the quantum channel is expressed as $\ket{\psi_\mathrm{hyb}} = \frac{1}{\sqrt{2}}(\ket{H}^{\otimes N}\ket{C_0} + \ket{V}^{\otimes N}\ket{C_1})$,
where $\ket{C_0}$ and $\ket{C_1}$ are the basis states for the carrier qubit.
We consider the three types of hybrid entangled states
\begin{align}
\label{eq:hyb_entanglement}
    \ket{\psi_{\mathrm{mc}}} &=  \frac{1}{\sqrt{2}}\left(\ket{H}^{\otimes N}\ket{\alpha}+\ket{V}^{\otimes N}\ket{-\alpha}\right), \nonumber\\
    \ket{\psi_\mathrm{mp}} &= \frac{1}{\sqrt{2}}\left(\ket{H}^{\otimes N}\ket{H}+\ket{V}^{\otimes N}\ket{V}\right), \nonumber\\
    \ket{\psi_\mathrm{ms}} &= \frac{1}{\sqrt{2}}\left(\ket{H}^{\otimes N}\ket{0} + \ket{V}^{\otimes N}\ket{1} \right),
\end{align}
where subscripts m, c, p and s denote multiphoton qubit, coherent-state qubit, PSP qubit, and VSP qubit, respectively.

We assume an asymmetric environment that the transmittance (reflectance) of every mode of the multiphoton qubit  is $t_M$ ($r_M$) and that of the carrier qubit is $t_C$ ($r_C$). 
Using Eq.~(\ref{eq:beamsplitter}), the shared hybrid entangled states are obtained as
\begin{align}\label{eq:mc_hyb}
    &\rho_{\mathrm{mc}}(t_M, t_C) = \frac{t_M^{2N}}{2}\Big\{ (\dyad{H})^{\otimes N}\otimes \dyad{t_C \alpha } \nonumber\\
    &~~~~~~~~~~~~~~~~~~+ (\dyad{V})^{\otimes N}\otimes \dyad{- t_C \alpha} \nonumber\\
    &~~~~~~~~~~~+ e^{-2 \alpha^2 r_C^2} \big[(\dyad{H}{V})^{\otimes N} \otimes \dyad{t_C \alpha}{-t_C \alpha} +\mathrm{H.c} \big] \Big\} \nonumber\\
    &~~~~~~~~~~~~~~~~~~+ (1-t_M^{2N})\rho^\mathrm{loss}_{\mathrm{mc}},
\end{align}
\begin{align}\label{eq:mp_hyb}
&\rho_\mathrm{mp}(t_M, t_C) = t_M^{2N} \Big\{t_C^{2} \dyad{\psi_{\mathrm{mp}}} +r_C^2 \big[(\dyad{H})^{\otimes N}\nonumber\\
&~~~~~~~~~~~~+ (\dyad{V})^{\otimes N} \big] \otimes \dyad{0} \Big\} +(1-t_M^{2N})\rho^\mathrm{loss}_{\mathrm{mp}},
\end{align}
and 
\begin{align}\label{eq:ms_hyb}
    &\rho_\mathrm{ms}(t_M, t_C) = \frac{t_M^{2N}}{2} \Big\{ (\dyad{H})^{\otimes N} \otimes \dyad{0}\nonumber\\
    &~~~~~~~~~~~~~~~~~~+(\dyad{V})^{\otimes N} \otimes (t_C^2 \dyad{1} + r_C^2 \dyad{0}) \nonumber\\
    &~~~~~~~~~~~~~~~~~~+ t_C \big[ (\dyad{H}{V})^{\otimes N} \otimes \dyad{0}{1} + \mathrm{H. c.} \big] \Big\}\nonumber\\
    &~~~~~~~~~~~~~~~~~~+ (1-t_M^{2N})\rho^\mathrm{loss}_\mathrm{ms},
\end{align}
where the loss terms $\rho^\mathrm{loss}$ represent the events where one or more photons are lost from the multiphoton qubit. 
Explicit expressions of the loss terms are
\begin{align*}
    \rho^\mathrm{loss}_{\mathrm{mc}}& = \frac{1}{2}\sum_{k=1}^{N}  (t_M^{2})^{N-k}(1-t_M^2)^{k} \sum_{\mathcal{P}\in \mathrm{Perm}(N,k)}\\
    &\Big\{\mathcal{P}\big[(\dyad{H})^{\otimes N-k}\otimes (\dyad{0})^{\otimes k}\big]\otimes \dyad{t_C \alpha}\\
    &+ \mathcal{P}\big[(\dyad{V})^{\otimes N-k}\otimes (\dyad{0})^{\otimes k}\big]\otimes \dyad{-t_C \alpha}\Big\},
\end{align*}
\begin{align*}
    &\rho^\mathrm{loss}_{\mathrm{mp}} =  \frac{1}{2}\sum_{k=1}^{N}  (t_M^{2})^{N-k}(1-t_M^2)^{k} \sum_{\mathcal{P}\in \mathrm{Perm}(N,k)}\\
    &~~\Big\{\mathcal{P}\big[(\dyad{H})^{\otimes N-k}\otimes (\dyad{0})^{\otimes k}\big]\otimes (t_C^2\dyad{H} + r_C^2\dyad{0})\\
    &+ \mathcal{P}\big[(\dyad{V})^{\otimes N-k}\otimes (\dyad{0})^{\otimes k}\big]\otimes (t_C^2\dyad{V} + r_C^2\dyad{0})\Big\},
\end{align*}
and 
\begin{align*}
    &\rho^\mathrm{loss}_{\mathrm{ms}} =  \frac{1}{2}\sum_{k=1}^{N}  (t_M^{2})^{N-k}(1-t_M^2)^{k} \sum_{\mathcal{P}\in \mathrm{Perm}(N,k)}\\
    &~~\Big\{\mathcal{P}\big[(\dyad{H})^{\otimes N-k}\otimes (\dyad{0})^{\otimes k}\big]\otimes (t_C^2\dyad{1} + r_C^2\dyad{0})\\
    &~~~~~~~~+ \mathcal{P}\big[(\dyad{V})^{\otimes N-k}\otimes (\dyad{0})^{\otimes k}\big]\otimes \dyad{0}\Big\}.
\end{align*}
All these terms  do not contain entanglement. This is attributed to the fact that when a photon from the multiphoton qubit is lost, the resulting multiphoton qubit effectively becomes completely dephased. 

\subsection{\label{subsec:entanglement}Amount of entanglement in hybrid entangled states}

In this subsection, we investigate the amount of entanglement contained in the hybrid entangled states. Entanglement in any bipartite mixed state can be measured by the negativity $\mathcal{N}(\rho)$ \cite{Vidal02}, which is defined as 
\begin{align}
    \mathcal{N}(\rho)\equiv\frac{\left\Vert\rho^{T_A}\right\Vert-1}{2} = \sum_{\lambda_i <0}|\lambda_i|
   \label{eq:Neg}
\end{align}
where $\rho^{T_A}$ is the partial transpose of $\rho$ with respect to subsystem $A$, $\left\Vert\cdot\right\Vert$ is the trace norm, and $\{\lambda_i \}$ is the set of eigenvalues of $\rho^{T_A}$. The negativity is an entanglement measure, i.e., it does not increase under any local operations and classical communications.

Using Eq.~(\ref{eq:Neg}), analytical expressions of the negativity of the hybrid entangled states can be obtained from Eqs.~(\ref{eq:mc_hyb}), (\ref{eq:mp_hyb}) and (\ref{eq:ms_hyb}). Although $|t_C \alpha\rangle$ and $|-t_C \alpha\rangle$ in  Eq.~(\ref{eq:mc_hyb}) are not orthogonal, they are two linear independent state vectors that can be treated in a two-dimensional Hilbert space as done in Ref.~\cite{Jeong01}.
Further, since the loss terms, $\rho^\mathrm{loss}$, are orthogonal to the remaining terms and contain no entanglement, we can consider only the remaining terms in a  $2 \otimes 2$ dimensional Hilbert space. 
 The degrees of negativity are then obtained as
\begin{align*}
    &\mathcal{N}(\rho_\mathrm{mc}) = \frac{t^{2N}_M} {4\sqrt{1-e^{-4t_C^2\alpha^2}}} \nonumber \\
    &\qquad \times \left[ \sqrt{ 1 - 2\left( 2e^{-4t_C^2\alpha^2} - 1 \right) e^{-2r_C^2\alpha^2} + e^{-4r_C^2\alpha^2} } \right. \nonumber \\
    &\qquad\qquad \left.+ e^{-2r_C^2\alpha^2} - 1 \right], \\
    &\mathcal{N}(\rho_\mathrm{mp}) = \mathcal{N}(\rho_\mathrm{ms}) = \frac{1}{2} t_M^{2N} t_C^2. 
\end{align*}
Here, the negativities of $\rho_\mathrm{mp}$ and $\rho_\mathrm{ms}$ are same, because entanglement disappears when at least one photon is definitely lost in both cases.

\begin{figure}[t]
\centering
\includegraphics[width=1\linewidth]{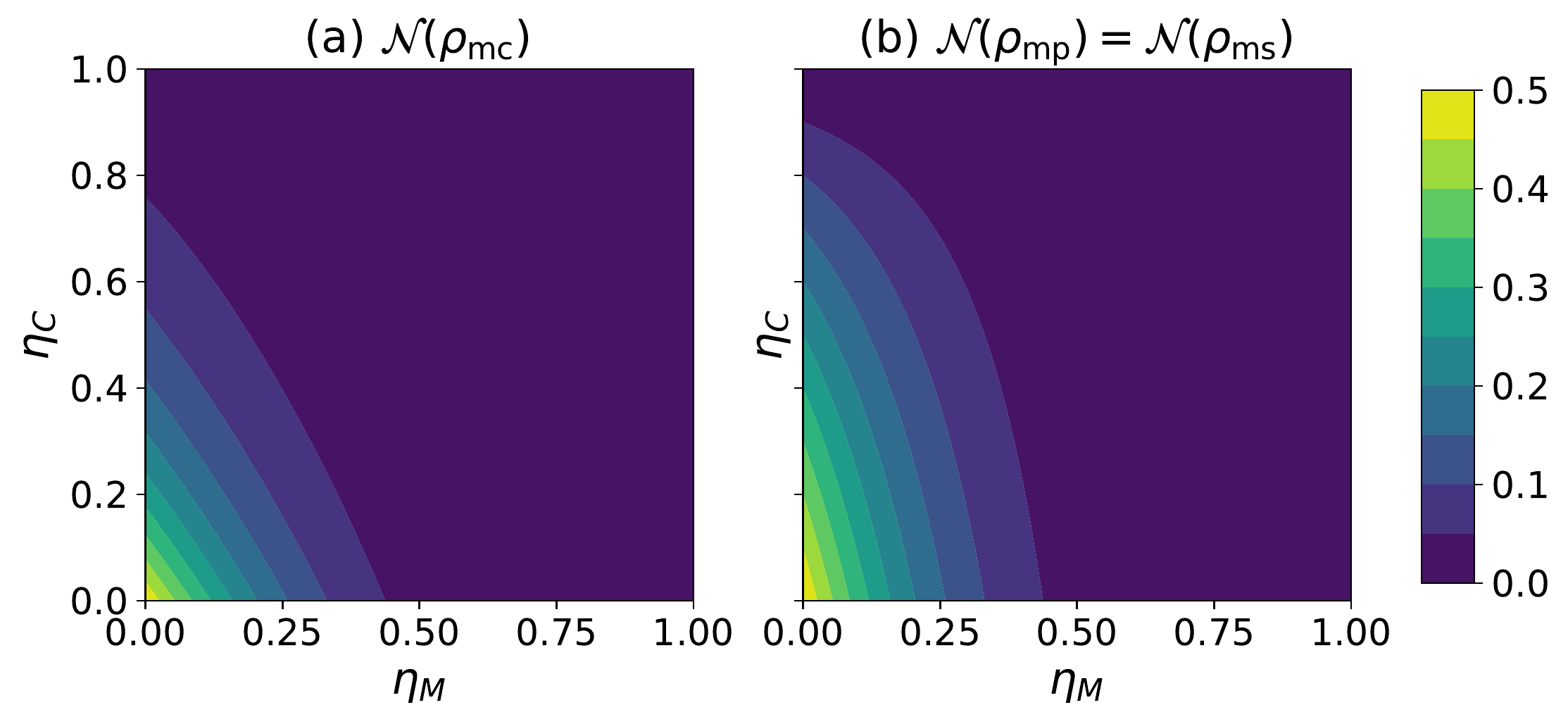}
\caption{
  Degrees of entanglement (negativity) against the photon-loss rate for the multiphoton qubit $\eta_M = 1-t_M^2$ and for the carrier qubit $\eta_C=1-t_C^2$ of hybrid entanglement between (a) the multiphoton qubit and the coherent-state qubit $\rho_{\mathrm{mc}}$, (b) the multiphoton qubit and the PSP qubit $\rho_{\mathrm{mp}}$, and the multiphoton qubit and the VSP qubit $\rho_{\mathrm{ms}}$. The number of photons $N$ for the multiphoton qubit is set to be $N=4$. The amplitude of the coherent-state qubit is chosen to be $\alpha = 1.2$.
}
\label{fig:negativity}
\end{figure}

Figure~\ref{fig:negativity} shows the dependence of the negativity on the photon loss rates of both the sides, $\eta_M = 1-t_M^2$ and $\eta_C = 1-t_C^2$. It is generally shown that the dependence is sharper for the loss rate, $\eta_M$,  of the multiphoton qubit  than that of the carrier qubit, $\eta_C$. This implies the desirable property that entanglement in the hybrid entangled state is more robust to the photon loss on the carrier qubit than that on the multiphoton qubit.

\subsection{\label{subsec:fidelity}Teleportation fidelities}

We now consider quantum teleportation with the hybrid entangled states $\rho_\mathrm{mc}$, $\rho_\mathrm{mp}$, and $\rho_\mathrm{ms}$ as the quantum channel. We employ the Bell-state measurement scheme for the multiphoton qubits proposed in Ref.~\cite{Lee15}. In the multiphoton qubit encoding, the Bell states are defined as
\begin{align*}
    \ket{B_{1,2}^N} & = \frac{1}{\sqrt2}\left(\ket{H}^{\otimes N}\ket{H}^{\otimes N}\pm\ket{V}^{\otimes N}\ket{V}^{\otimes N}\right), \nonumber\\
    \ket{B_{3,4}^N} & = \frac{1}{\sqrt2}\left(\ket{H}^{\otimes N}\ket{V}^{\otimes N}\pm\ket{V}^{\otimes N}\ket{H}^{\otimes N}\right),
\end{align*}
where $\pm$ is chosen in the same order of the two number labels of $\ket{B_i^N}$ in each line. Using only linear optics and on-off photodetectors, $\ket{B_2^N}$ and $\ket{B_4^N}$ are identified  unambiguously while $\ket{B_1^N}$ and $\ket{B_3^N}$ with probability $1-1/2^{N-1}$  \cite{Lee15}. 

When one or more photons are lost from the muliphoton qubit in hybrid entangled states, there is a chance that $\ket{B_i^K}$ with $K<N$ is detected. However, although we accept these events as success,  the fidelity is not improved. We pointed out earlier that $\rho^\mathrm{loss}$ does not contain entanglement due to the dephasing induced by photon loss. The teleportation fidelity between the input and output qubits cannot then exceed the classical limit, which we will discuss further at the end of this section. We thus take only the detection of $N$-photon Bell states as the successful events.

Similarly to the standard teleportation scheme, a sender jointly measures the input state $\ket{\psi_\mathrm{in}}=a \ket{H}^{\otimes N} + b \ket{V}^{\otimes N}$ and the multiphoton-qubit part of the hybrid entangled states. After the Bell-state measurement with  outcome $i$, the input state $\ket{\psi_\mathrm{in}}$ and the hybird entangled state under photon loss, $\rho_\mathrm{hyb}(t_M, t_C)$, are projected to
\begin{align}\label{eq:postmeasurment}
    \rho_{\mathrm{out},i}(t_M, t_C) = \frac{\bra{B_i^N} (\dyad{\psi_\mathrm{in}}\otimes \rho_\mathrm{hyb}(t_M, t_C) )\ket{B_i^N}}{\tr [\dyad{B_i^N} (\psi_\mathrm{in} \otimes \rho_\mathrm{hyb}(t_M, t_C))]}.
\end{align}
With the heralded measurement outcome $i$, the receiver may recover the state $\rho_\mathrm{out} = \rho_\mathrm{out, 1}$ by a proper local unitary based on the outcome $i$. 

Before proceeding further, we point out  that the output state does not depend on  loss $\eta_M$ on the multiphoton-qubit part. The hybrid entangled state can be represented as $\rho_\mathrm{hyb}(t_M, t_C) = t_M^{2N} \sigma_\mathrm{hyb}(t_C)+ \rho^\mathrm{loss}$, where $\sigma_\mathrm{hyb}(t_C)$ corresponds to the state when  no  photon is lost from the multiphoton qubit. The facter $t_M^{2N}$ indicates that this event happens with a probability of $t_M^{2N} = (1-\eta_M)^N$. Since the loss term $\rho^\mathrm{loss}$ is orthogonal to the qubit basis $\{\ket{H}^{\otimes N}, \ket{V}^{\otimes N}\}$, only $\sigma_\mathrm{hyb}(t_C)$ remains after the projection on $\ket{B_N^i}$. The factor $t_M^{2N}$ in both the numerator and the denominator of Eq.~(\ref{eq:postmeasurment}) then cancels out. Thus, $\rho_\mathrm{out}(t_M, t_C)$ is independent of $t_M$ so that it can be represented as $\rho_\mathrm{out}(t_C)$.

We set the target state of the teleportation to be $\ket{\psi_\mathrm{t}} = a\ket{C_0}+b\ket{C_1}$. The quantum fidelity between the output state $\rho_\mathrm{out}$ and the target state $\ket{\psi_\mathrm{t}}$ is defined as 
\begin{align*}
    F(t_C)= \expval{\rho_\mathrm{out}(t_C)}{\psi_\mathrm{t}}.
\end{align*}

Now, we examine the candidates of the carrier qubit. First, we consider quantum teleportation from a multiphoton qubit to a coherent-state qubit. When $\ket{B_1^N}$ is detected, we can express the output qubits after Bell-state measurement by 
\begin{align*}
\rho_{\mathrm{out}, 1}^\mathrm{m \rightarrow c} =& M_+ \big[\abs{a}^2 \dyad{t_C \alpha}+\abs{b}^2\dyad{-t_C \alpha} \nonumber\\
&+e^{-2\alpha^2 r_C^2}(ab^* \dyad{t_C\alpha}{-t_C\alpha}+\mathrm{H.c.})\big],
\end{align*}
where $M_+=\big[ 1+e^{-2\alpha^2} (a b^*+a^* b) \big]^{-1}$. When $\ket{B_3^N}$ is detected, the output qubit undergoes a bit flip as $\rho_\mathrm{out, 3}^{\mathrm m \rightarrow c} = X_c \rho_\mathrm{out, 1}^{\mathrm m \rightarrow c} X_c^{\dagger}$ with $X_c : \ket{\pm t_C \alpha} \rightarrow \ket{\mp t_C \alpha}$. This effect can be corrected by applying a $\pi$-phase shifter. However, when $\ket{B_2^N}$ is detected, the output qubit becomes
\begin{align*}
    \rho_{\mathrm{out}, 2}^\mathrm{m \rightarrow c}=& M_- \big[\abs{a}^2 \dyad{t_C \alpha}+\abs{b}^2\dyad{-t_C \alpha} \nonumber\\
    &-e^{-2\alpha^2 r_C^2}(ab^* \dyad{t_C\alpha}{-t_C\alpha}+\mathrm{H.c.})\big],    
\end{align*}
which cannot be corrected to $\rho_\mathrm{out, 1}^\mathrm{m \rightarrow c}$  by applying a unitary operation
because of   the nonorthogonality of the coherent-state qubit basis. 
In other words, the required operation $Z_c : \ket{\pm t_C \alpha} \rightarrow \pm \ket{\pm t_C \alpha}$
cannot be performed in a fully deterministic way.
 There are, however, approximate methods to perform the required $Z_c$ operation using the displacement operation \cite{Jeong01, Jeong02} or the gate teleportation protocol \cite{Ralph03}.
  We also note that the transformation of $\rho_{\mathrm{out}, 4}^\mathrm{m \rightarrow c} \rightarrow \rho_{\mathrm{out}, 2}^\mathrm{m \rightarrow c}$ can be carried out by the $X_c$ gate.  Therefore, the output qubit is  one of the non-exchangable states, $\rho_{\mathrm{out}, 1}^\mathrm{m \rightarrow c}$ or $\rho_{\mathrm{out}, 2}^\mathrm{m \rightarrow c}$. We denote these two states as $\rho_\mathrm{out, \pm}^\mathrm{m \rightarrow c}$.  Nevertheless, the measurement outcome $i$ heralds the transformation of the output states. Thus, the output qubit has the quantum information of the initial qubit. 

Given the  transmittance $t_C$, we take the dynamical qubit basis $\qty{\ket{t_C \alpha}, \ket{-t_C \alpha}}$ as the output qubit basis. As an analogy of the input state, we set the two target states as
\begin{align*}
\ket{ \psi_\mathrm{t, \pm}^{m \rightarrow c} } = N_\pm \qty( a\ket{t_C \alpha} \pm b\ket{ - t_C \alpha} ),
\end{align*}
where $N_\pm =\qty{1 \pm (a b^* + a^* b ) \exp (-2t^2 \alpha^2)}^{-1/2}$ are the normalization constants. Then, we obtain the fidelity between $\rho_{\mathrm{out}, \pm}^\mathrm{m \rightarrow c}$ and $\ket{\psi_\mathrm{t, \pm}^\mathrm{m \rightarrow c}}$ respectively:
\begin{align*}
    F^{m \rightarrow c}_\pm &(t_C ; a, b) = \ev{\rho_{\mathrm{out}, \pm}^{m \rightarrow c}}{\psi_\mathrm{t, \pm}^\mathrm{m \rightarrow c}} \nonumber\\
    =& M_\pm N_\pm^2 \big[ \abs{a}^2 \abs{a\pm bS}^2 + \abs{b}^2 \abs{aS \pm b}^2 \nonumber\\
    &\pm 2 e^{-2\alpha^2 r_C^2} \mathrm{Re}\left[ ab^* (a^* \pm b^* S) (a S \pm b)\right] \big],
\end{align*}
where $S=\bra{t_C\alpha}\ket{-t_C\alpha}=e^{-2t_C^2 \alpha^2}$ is the overlap between the output coherent-state qubit basis states. We now compute the average fidelity over all input states. We use a parametrization $a = \cos(\theta/2)\exp(i \phi /2)$ and $b = \sin(\theta/2) \exp(-i \phi /2)$ with uniformly random sampling on the Bloch sphere. Note that $F^{m \rightarrow c}_+(t_C ; a, b) = F^{m \rightarrow c}_-(t_C ; a, -b)$, so the average fidelities of both cases are equal. Finally, we get the following integration:

\begin{align}
\label{eq:CS_ave_fid}
    &F_\mathrm{ave}^\mathrm{m \rightarrow c}(t_C) = \left< F^{m \rightarrow c}_\pm(t_C ; a, b)\right>_{\theta, \phi} \nonumber\\
    &~~~~~~~~~~~~= \frac{1}{4 \pi } \int_0^\pi d \theta \sin \theta \int_0^{2\pi} d\phi F^{m \rightarrow c}_\pm(t_C; \theta, \phi).
\end{align}

\begin{figure}
    \centering
    \includegraphics[width=1\linewidth]{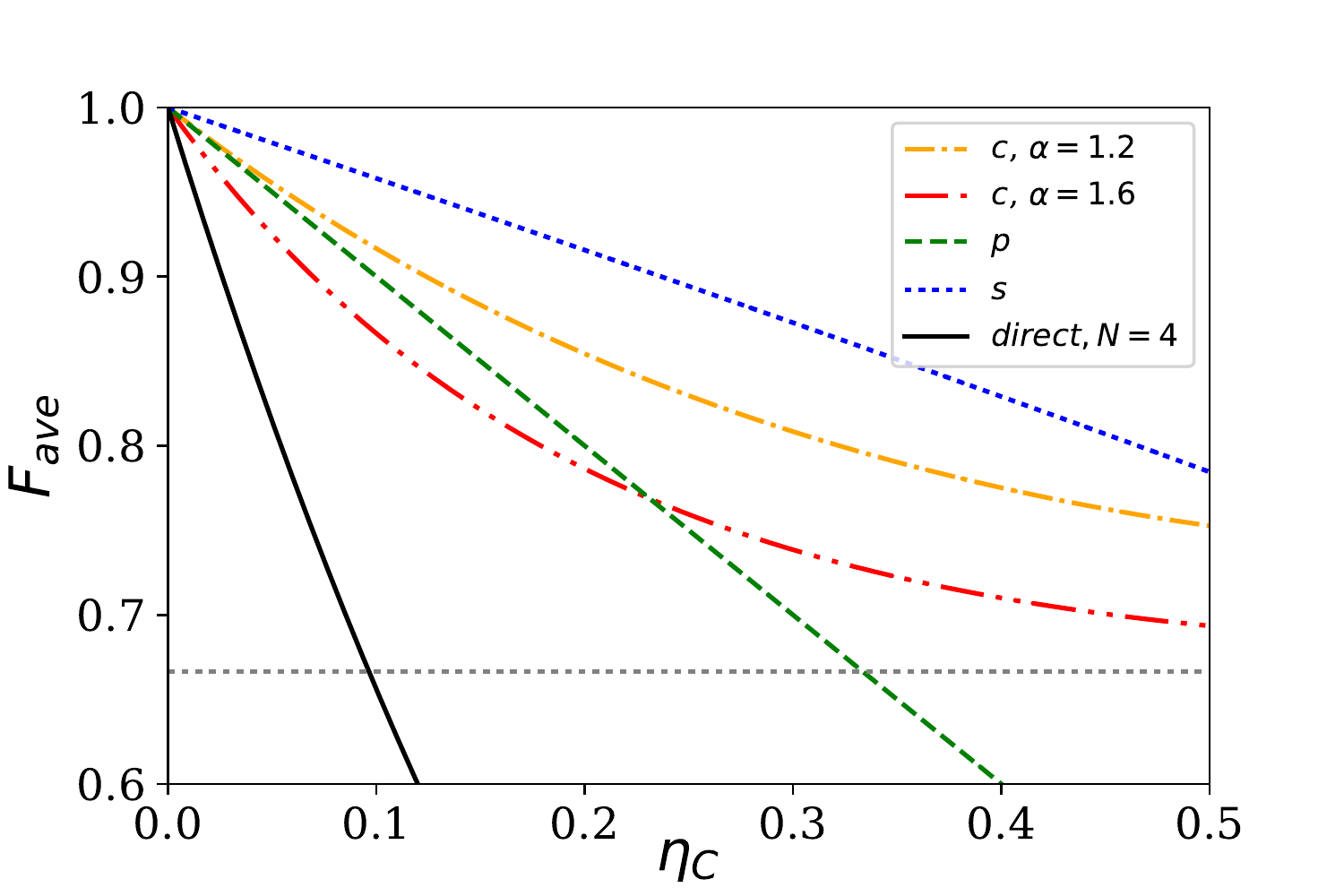}
\caption{Average fidelities of direct transmission with $N=4$ (black solid) and hybrid archtectures with different carrier qubits: coherent state qubits (denoted by c) for $\alpha = 1.2$ (yellow dot-dashed) and $\alpha = 1.6$ (red dot-dot-dashed), a PSP qubit (p, green dashed), and a VSP qubit (s, blue dotted) against the photon-lose rate for the carrier-qubit part $\eta_C = 1-t_C^2$. The gray horizontal dotted line is the classical limit $F_\mathrm{cl} = 2/3$.}
\label{fig:fid_comparison}
\end{figure}

The analytic expression of this integration is given in Ref.~\cite{Park12} but is too lengthy to present here. We show the average fidelity varying  amplitude $\alpha$ of the coherent-state qubit  in Fig.~\ref{fig:fid_comparison} (a). The plot shows that as the mean photon number $\alpha^2$ is smaller, the average fidelity approaches the unity. However, a small value of $\alpha$ makes the overlap between $\ket{\pm\alpha}$ large so that its ability for quantum information processing (for example, the probability to perform $Z_c$ gate) becomes low. 

In the case of the quatum teleporatation from multiphoton qubit to PSP qubit, we use the hybrid entangled state in Eq.~(\ref{eq:mp_hyb}). Since all single-qubit operations can be implemented in linear optics \cite{Knill01, Dodd03}, we set the unique target state: $\ket{\psi_\mathrm{t}^\mathrm{m \rightarrow p}} = a \ket{H} + b\ket{V} $. When a Bell state $\ket{B_1^N}$ is detected, the output state is 
\begin{align*}
\rho_{\mathrm{out}, 1}^\mathrm{m \rightarrow p} &= t_C^{2}\big(\abs{a}^2 \dyad{H} + \abs{b}^2 \dyad{V} \nonumber\\
&\quad+ ab^*\dyad{H}{V} + a^* b\dyad{V}{H}\big) + r_C^2\dyad{0} \nonumber\\
&= t_C^2 \dyad{\psi_\mathrm{t}^\mathrm{m \rightarrow p}} + r_C^2 \dyad{0}.
\end{align*}
When the other Bell states are detected, after receiving the measurement outcome, the receiver can recover the target state by a proper single-qubit unitary operation. The  fidelity is then readily obtained as

\begin{align*}
    F^\mathrm{m \rightarrow p}(t_C) = t_C^2.
\end{align*}

The last case is teleportation from a multiphoton qubit to a VSP qubit with the entangled state in Eq. (\ref{eq:ms_hyb}). In this case of the VSP qubit, the situation is similar to the case of the coherent-state qubit. While the Z operation is deterministic in linear optics, the $X$ operation, $X:\ket{0} \leftrightarrow \ket{1}$, is probabilistic \cite{Lund02}. Therefore, we  distinguish the output qubit of $\ket{B_1^N}$ detection, denoting $\rho_\mathrm{out, +}^\mathrm{m\rightarrow s}$, from $\ket{B_2^N}$, denoting $\rho_\mathrm{out, -}^\mathrm{m\rightarrow s}$. The output qubit when $\ket{B_1^N}$ is detected is obtained  similarly as
\begin{align*}
\rho_\mathrm{out, +}^\mathrm{m \rightarrow s} &= (\abs{a}^2 + \abs{b}^2r_C^2) \dyad{0}+ \abs{b}^2 t_C^2 \dyad{1} \nonumber\\
&\quad+ (a b^* t_C \dyad{0}{1} + \mathrm{H. c.}). 
\end{align*}
We then obtain the input-dependent fidelity as
\begin{align*}
F^\mathrm{m \rightarrow s} (t_C) = \abs{a}^4  + \abs{a}^2 \abs{b}^2(1+ t_C) + \abs{b}^4 t_C^2.
\end{align*}
In this case, the average fildelity has a simple analytic expression: 
\begin{align*}
F_\mathrm{ave}^\mathrm{m \rightarrow s}(t_C) = \frac{1}{3} t_C^2 + \frac{1}{6} t_C + \frac{1}{2}.
\end{align*}

We need to consider the classical fidelity $F_\mathrm{cl}$ that is defined as the maximum average fidelity obtained by teleportation protocol without entanglement. It is well known that $F_\mathrm{cl} = 2/3$ for a qubit with an orthonormal basis \cite{Massar95}. If we use the coherent-state qubit of $\ket{\pm\alpha}$ as the carrier qubit, however, $F_\mathrm{cl}^\mathrm{m \rightarrow c}$ is
\begin{align*}
    F_\mathrm{cl}^\mathrm{m \rightarrow c}(t_C) = \frac{S + 3S^2 - (S^4-1)}{4S^3} \sinh ^{-1} \qty[\frac{S}{\sqrt{1-S^2}}],
\end{align*}
where $S=\braket{-t_C\alpha}{t_C \alpha}=e^{-2t_C^2 \alpha^2}$ \cite{Park12}. In this case, the classical limit becomes larger $2/3$ due to the nonorthogonality $S$. Of course, $F_\mathrm{cl}^\mathrm{m \rightarrow c}$ converges to $2/3$ as $S \rightarrow 0$. In Fig.~\ref{fig:fid_comparison}, the average fidelity $F_\mathrm{cl}^\mathrm{m\rightarrow c}$ is approximately $2/3$ for the area of $\alpha \geq 1.2$ and $\eta \leq 0.5$.

In Fig.~\ref{fig:fid_comparison}, we present  the average fidelities between the output qubit and the target state against the photon-loss rate $\eta_C$ for the different types of the carrier qubit. For the coherent-state qubit, we choose amplitudes of $\alpha = 1.2$ and $1.6$, which are approximately the minimum and optimal amplitudes, respectively, for the fault-tolerant quantum computing using the 7-qubit Steane code \cite{Lund08}. Obvious, better fidelities over the direct transmission can be obtained using the teleportation protocol. Among the carrier qubits, the VSP qubit is better than the PSP qubit. The reason for this can be understood as follows. When photon loss occurs, the PSP qubit gets out of the original qubit space because of the vacuum portion. However,  the VSP qubit remains in the original qubit space even under the photon loss.

The comparison between the coherent-state qubit and  the other types of qubits depends on amplitude $\alpha$. With small values of $\alpha$, the coherent-state qubit shows higher average fidelity than the others. We numerically obtain that, when $\alpha<1.23$ ($\alpha<0.78$), the average fidelity of the corresponding coherent-state qubit is higher  than that of the PSP qubit (the VSP qubit) for any rates of photon loss. However, one should note that the overlap between two coherent states $\ket{\pm \alpha}$ is $\braket{\alpha}{-\alpha} = \mathrm{exp} (-2\alpha^2) \approx 0.0485$ ($0.296$) for $\alpha = 1.23$ ($0.78$), which could be a negative factor depending on the task to perform.

\begin{table}[b]\label{tab:loss_limit}
  \begin{center}
    \caption{Maximum photon-loss rates  for the carrier qubit,  $\eta_C$, required to reach the fidelity of 99.9\%, 99\%, and 90\% with the coherent-state (CS) qubit, the PSP qubit, and the VSP qubit. The direct transmission (DT) of the multiphoton qubit with the photon number $N=4$ is given for  comparison under the same photon-loss rate.}
    \label{tab:lim_loss}
    \begin{tabular}{p{1.3cm} *{5}{>{\centering}p{1cm}} c}
      \hline
      \hline
      \multirow{2}{*}{$F$} & \multirow{2}{*}{DT} & \multicolumn{2}{c}{CS} & \multirow{2}{*}{PSP} & \multirow{2}{*}{VSP}& \\ \cline{3-4}
      & &1.2 & 1.6 & & & \\
      \hline
      99.9\% & 0.025 & 0.10 & 0.059 & 0.10 & 0.24 & \multirow{3}{*}{($\times 10^{-2}$)} \\
      99\% & 0.25 & 1.1 & 0.59 & 1.0 & 2.4 &\\
      90\% & 2.6 & 12 & 7.0 & 10 & 24 &\\
      \hline
      \hline
    \end{tabular}
  \end{center}
\end{table}

All-optical quantum computing schemes have tolerable limits of photon loss rates for  fault tolerance \cite{Dawson06, Lund08, Herrera10, Lee13, Li15, Wehner18}. In Table \ref{tab:lim_loss}, we summarize the maximum photon-loss rates for the carrier qubit, $\eta_C$, which can be tolerated while preserving the fidelity to be 99.9\%, 99\%, and 90\% within our hybrid architectures. In this high fidelity regime, the VSP qubit tolerates approximately 10 times larger photon loss than the direct transmission.

\subsection{\label{subsec:sucprob}Success probabilities}

\begin{figure}
\includegraphics[width=\linewidth]{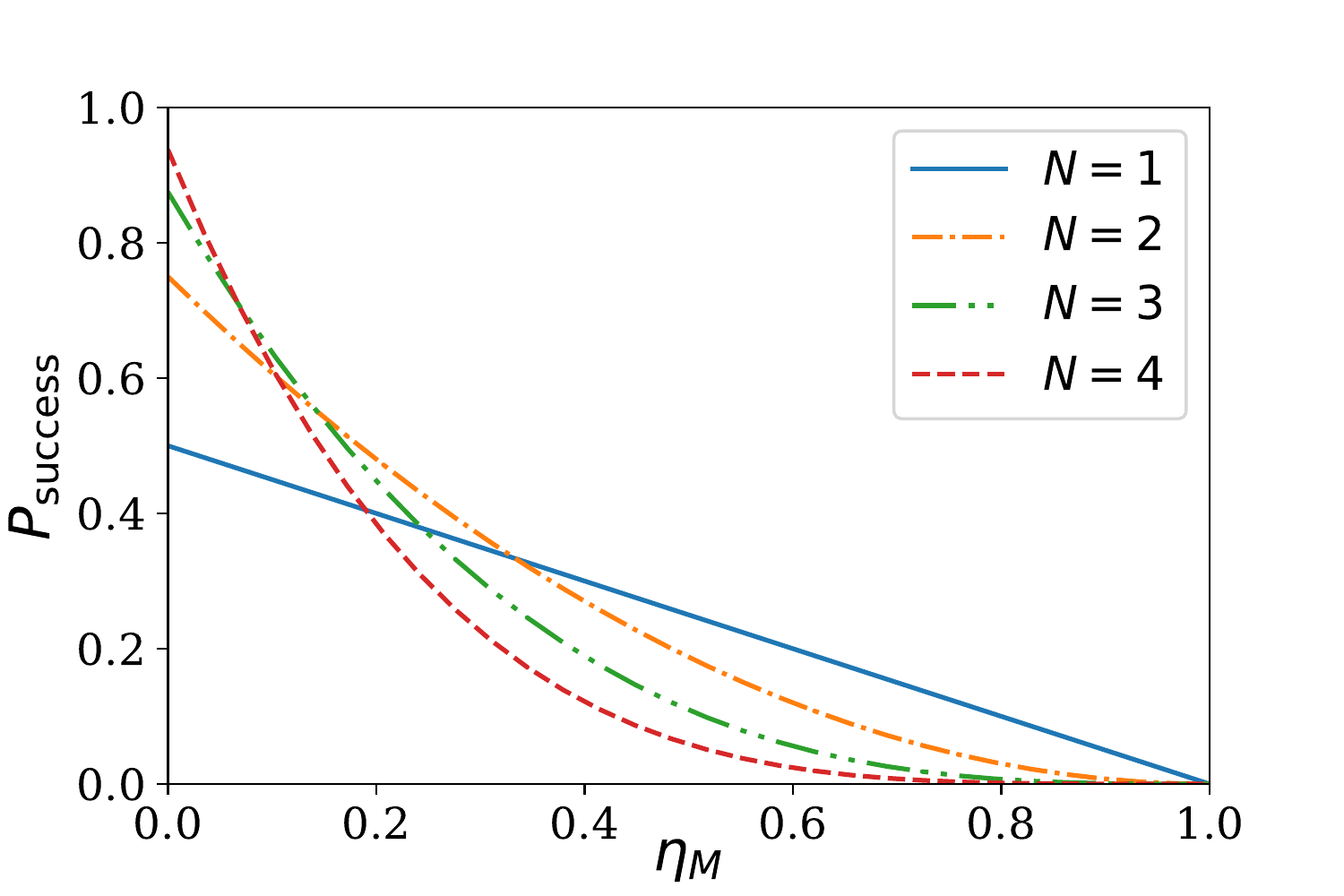}
\caption{Success probability $P_\mathrm{success}$ of the multiphoton Bell-state measurement against the photon-loss rate for the multiphoton qubit $\eta_M = 1-t_M^2$ for photon numbers of $N =$1, 2, 3, and 4.}
\label{fig:suc_comparision}
\end{figure}

Only when the input qubits are in the logical qubit basis and the identification between $\ket{B_1^N}$ and $\ket{B_3^N}$ is successful, the Bell-state measurement successes. Let us denote $q_i$ as the probability of the successful identification of $\ket{B_i^N}$ when $\ket{B_i^N}$ is given. This $q_i$ varies according to the Bell-state measurement scheme, and we follows the Bell-state measurement scheme of multiphoton qubit that $q_i= 1-1/2^{N-1}$ for $i = \mathrm{odd}$ and $q_i = 1$ for $i = \mathrm{even}$ \cite{Lee15}. The success probability of the teleportation with the hybrid entangled state $\rho_\mathrm{hyb}$ is then given as 
\begin{align}
    P = \sum_i q_i \tr \Big[\ket{B_i^N}_\mathrm{SS'}\bra{B_i^N}\qty(\ket{\psi_\mathrm{in}}_\mathrm{S} \bra{\psi_\mathrm{in}}\otimes(\rho_\mathrm{hyb})_\mathrm{S'R})\Big],
\label{eq:prob-hyb}
\end{align}
where S and S' represents the sender's modes and R does the receiver's mode. Note that the success probability $P$ does not depend on $t_C$ since \begin{align*}
&\tr \Big[\ket{B_i^N}_\mathrm{SS'}\bra{B_i^N}\qty(\ket{\psi_\mathrm{in}}_\mathrm{S} \bra{\psi_\mathrm{in}}\otimes(\rho_\mathrm{hyb})_\mathrm{S'R})\Big] \\
&= \tr_{\rm SS'} \Big[\ket{B_i^N}_\mathrm{SS'}\bra{B_i^N}\qty(\ket{\psi_\mathrm{in}}_\mathrm{S} \bra{\psi_\mathrm{in}}\otimes (\tr_{\mathrm R}\rho_\mathrm{hyb})_\mathrm{S'})\Big] 
\end{align*}
and $\tr_\mathrm{R}\rho_\mathrm{hyb}(t_M, t_C) = \tr_\mathrm{R}(\Phi_{t_M}\otimes I)(\dyad{\psi_\mathrm{hyb}})$ 
from the trace-preserving property of $\Phi_{t}$
where R represents the receiver's mode and $\Phi_{t_M}$ is the quantum channel of photon loss with transmittance $t_M$. 

Now, we examine the success probability for each carrier qubit. For the case of  teleportation to a coherent-state qubit with $\rho_\mathrm{mc}$ in Eq. (\ref{eq:mc_hyb}), the success probability of the teleportation, $P^\mathrm{m \rightarrow c}$, is obtained using Eq.~(\ref{eq:prob-hyb}) as
\begin{align*}
    &P^\mathrm{m \rightarrow c} (t_M, N; a, b) \nonumber\\
    &~~~~~~~~~~~~=  t_M^{2N}\qty[ \qty(1- \frac{1}{2^N}) - \frac{e^{-2\alpha^2}}{2^N}\qty(a b^* + a^* b)].
\end{align*}
The last term is from the nonorthogonality of the coherent state qubit. We also obtain the averaged success probability $P_\mathrm{ave}^\mathrm{m\rightarrow c}$ by averaging $P^{m \rightarrow c} (t_M, N; a, b)$ on all possible input state with the same parametrization of Eq. (\ref{eq:CS_ave_fid}) as
\begin{align*}
    P_\mathrm{ave}^\mathrm{m \rightarrow c} (t_M, N) = t_M^{2N}\qty(1- \frac{1}{2^N}).
\end{align*}
For the discrete variable qubits, we have $\tr_{\mathrm R} (\Phi_{t_M}\otimes I)\dyad{\psi_\mathrm{hyb}} = \Phi_{t_M}(I/2)$. Therefore, without dependence on the carrier qubit, we obtain 
\begin{align}\label{eq:Suc_prob}
    P(t_M, N) = t_M ^{2N} \left( 1- \frac{1}{2^N} \right).
\end{align}

In Fig.~\ref{fig:suc_comparision}, we plot the success probability $P(t_M, N)$ as a function of the photon loss rate for the multiphoton qubit, $\eta_M$, by changing the photon number $N$ of the multiphoton qubit. The success probability in Eq. (\ref{eq:Suc_prob}) shows an interesting feature: while the success probability of Bell-state measurement increases with $N$ for $t_M = 1$, if $t_M$ is less than 1, larger $N$ rather makes the success probability lower. This supports the general belief that a ``macroscopic object'' is fragile under loss if we regard the larger $N$ means the qubit is more ``macroscopic'' \cite{Frowis18}. It is straightforward to obtain the optimal number of photons per a multiphoton qubit, $N_\mathrm{opt} = \lfloor \log_2 (1+1/\eta_M) \rfloor$, that maximizes the success probability $P(t_M, N)$. 

\section{\label{sec:gen}Generation of  multiphoton hybrid entangled states}

In this section, we discuss how to generate the required hybrid entangled states $\ket{\psi_\mathrm{mc}}$, $\ket{\psi_\mathrm{mp}}$ and $\ket{\psi_\mathrm{ms}}$ in Eq.~(\ref{eq:hyb_entanglement}). 
We may start with a GHZ state of PSP qubits:
$ \ket{\mathrm{GHZ}(N)} =( \ket{H}^{\otimes N} + \ket{V}^{\otimes N} )/ \sqrt{2}$ .
It is then clear that $\ket{\psi_\mathrm{mp}} = (\ket{H}^{\otimes N}\ket{H}+\ket{V}^{\otimes N}\ket{V})/{\sqrt{2}}$ is simply a GHZ state with $N+1$ modes $\ket{\mathrm{GHZ}(N+1)}$. In addition to a GHZ state of $N+1$ photons $\ket{\mathrm{GHZ}(N+1)}$, we need to find out methods to convert one of the polarization qubits in the GHZ state to the desired carrier qubit by a conversion gate $V = \dyad{C_0}{H} + \dyad{C_1}{V}$. In this way, desired hybrid entangled states may be obtained.

There are a number of proposals for the generation of the GHZ state.
   A linear optical setup, called the (Type-I) fusion gate, is designed to fuse $\ket{\mathrm{GHZ}(N)}$ and $\ket{\mathrm{GHZ}(2)}$ to generate $\ket{\mathrm{GHZ}(N+1)}$ with a probability of $50\%$ \cite{Browne05}. 
In a similar method (Supplementary Material of Ref.~\cite{Varnava08}), 6 single photons are fused by the fusion gate followed by a Bell-state projection to generate $\ket{\mathrm{GHZ}(3)}$. The Bell-state measurements on copies of $\ket{\mathrm{GHZ}(3)}$ also provide a probabilistic method to generate a GHZ state with an arbitrary high photon number \cite{Varnava08, Li15}. Using the Bell-state measurements, this method is made robust to photon loss \cite{Varnava08}. Alternatively, a method  based on a nonlinear interaction, called coherent photon conversion, was proposed  to implement a deterministic photon-doubling gate $\dyad{HH}{H} + \dyad{VV}{V}$ \cite{Langford11}.
 So far, the multiphoton GHZ-type entanglement has been experimentally observed with postselection in most experiments  (for example, \cite{Zhang16, Zhang15, Huang11, Wang16, Zhong18}), which cannot be used as a teleportation channel. Nevertheless, a direct generation of a three-photon GHZ state was experimentally performed \cite{Hamel14}.

There are several methods to convert one PSP qubit to the VSP qubit~\cite{Ralph05, Fiurasek17, Drahi19}.
The conversion gate $V_\mathrm{p\rightarrow s}=\dyad{0}{H}+\dyad{1}{V}$ was experimentally demonstrated using the teleportation protocol and post-selection~\cite{Drahi19}.

In Ref.~\cite{Kwon15}, the authors suggested a method for conversion operation $V_\mathrm{p\rightarrow c} = \dyad{\alpha}{H}+\dyad{-\alpha}{V}$ using passive linear elements, single-photon detectors 
and a superposition of coherent states.
This scheme allows the conversion $V_\mathrm{p \rightarrow c} = \dyad{\alpha}{H}+\dyad{-\alpha}{V}$ using a superposition of coherent states with an amplitude slightly larger than $\alpha$. We note that a superposition of coherent states with amplitude $\alpha \approx 1.85$ in a traveling field was recently generated \cite{Sychev17}.

Experimental attempts to perform aforementioned proposals to generate multiphoton hybrid entangled states with high fidelities would be challenging due to effects of inefficient detectors, photon loss and other noisy effects.
It is, however, beyond the scope of this work to investigate and analyze those details under realistic conditions.

\section{Remarks}

It is important to identify efficient qubit encoding for a given quantum information task such as quantum communication and computation. 
The multiphoton encoding enables one to perform a nearly deterministic Bell-state measurement, which is a remarkable advantage for quantum communication and computation. 
However, a multiphoton qubit is vulnerable to photon loss and this is a formidable obstacle particularly against long-distance quantum communication. 
In order to overcome this problem, 
we have suggested a teleportation scheme via hybrid entanglement between a multiphoton qubit and another type of optical qubit serving as a loss-tolerant carrier. 
In our scheme, only the loss-tolerant carrier qubit is sent through a lossy environment, where the coherent-state qubit, the PSP qubit, and the VSP qubit are considered as the loss tolerant carrier qubit. 

We have found that the average fidelities of the teleportation with the considered hybrid entangled states are better than the direct transmission.
The VSP qubit in hybrid entanglement serves as the best carrier showing about 10 times better tolerance on the photon-loss rate than the direct transmission of the multiphoton qubit  for the fidelity larger than 0.9. Our numerical analysis further shows that the coherent-state qubit shows higher average fidelity than the others with small values of $\alpha$. When $\alpha<1.23$ ($\alpha<0.78$), the average fidelity of the corresponding coherent-state qubit is higher  than that of the PSP qubit (the VSP qubit) for any rates of photon loss.
These results would be useful information when choosing the proper carrier qubit depending on the quantum tasks under consideration.
We have also investigated the average success probability of the teleportation. It was shown that the success probability depends only on the loss of the multiphoton-qubit part. Although the Bell-state measurement scheme of the multiphoton qubit is nearly deterministic without loss, the photon loss limits the maximum success probability.
Our work may be useful for the optical realization of long-distance quantum information processing by exploring hybrid architectures of optical networks.

\section{Acknowledgement}
This work was supported by the National Research Foundation of Korea (NRF) through grants funded by the the Ministry of Science and ICT (Grants No. NRF-2019M3E4A1080074 and NRF-2020R1A2C1008609). 
S.C. was supported by NRF Grant funded by Korean Government (NRF-2016H1A2A1908381-Global Ph.D. Fellowship Program)

%


\end{document}